\documentclass[12pts]{article}
\usepackage{setspace}
\usepackage{graphicx}
\usepackage{amsthm, amsmath, mathrsfs}
\usepackage{hyperref}
\usepackage{multirow}
\usepackage{caption}
\usepackage{natbib}
\usepackage[]{algorithm2e}
\usepackage[letterpaper, margin=1.5in]{geometry}
\usepackage[FIGTOPCAP]{subfigure}
\usepackage{authblk}
\usepackage{numprint}
\usepackage{enumitem}
\usepackage[section]{placeins}

\usepackage{cleveref}
\newtheorem{prop}{Proposition}
\crefname{prop}{proposition}{propositions}
\crefname{algo}{algorithm}{algorithms}

\begin{document}

\title{Frictional Unemployment on Labor Flow Networks\thanks{The authors would like to acknowledge the useful comments from Ernesto Carrela, Andrew Elliot, Doyne Farmer, Daniel Fricke, Austin Gerig, Basile Grassi, Matthew Jackson, Dietmar Maringer, Ulrich Matter, Jos\'{e} Javier Ramasco, Felix Reed-Tsochas, Alois Schuste, Robert Shimer, Margaret Stevens, Meri Obstbaum and three anonymous reviewers. The authors also thank the participants of the CABDyN Complexity Centre Seminars, the University of Basel Economics Seminars, the INET Seminars, and the Nuffield College Economics postdoc seminars. This work was supported under the INET at Oxford grant INET12-9001 and by the Oxford Martin School grant LC1213-006.}
}

\author[1,2]{Robert L. Axtell}
\author[3,4,5]{Omar A. Guerrero}
\author[1]{Eduardo L\'opez}
\affil[1]{Department of Computational and Data Sciences, George Mason University}
\affil[2]{Santa Fe Institute}
\affil[3]{Department of Economics, University College London}
\affil[4]{The Alan Turing Institute}
\affil[5]{Department of Computer Science, Aalto University}
\renewcommand\Authands{, and }

\date{}

\maketitle

\begin{abstract}
We develop an alternative theory to the aggregate matching function in which workers search for jobs through a network of firms: the labor flow network. The lack of an edge between two companies indicates the impossibility of labor flows between them due to high frictions. In equilibrium, firms' hiring behavior correlates through the network, generating highly disaggregated local unemployment. Hence, aggregation depends on the topology of the network in non-trivial ways. This theory provides new micro-foundations for the Beveridge curve, wage dispersion, and the employer-size premium. We apply our model to employer-employee matched records and find that network topologies with Pareto-distributed connections cause disproportionately large changes on aggregate unemployment under high labor supply elasticity.
\end{abstract}

\doublespacing
\newpage
\section{Introduction}

Unemployment is a fundamental economic problem resulting from several distinct social mechanisms. These include people becoming separated from their jobs and searching for new positions; firms opening vacancies and searching for new workers; and recruiters finding job seekers throughout the labor market. Due to the complexity involved in simultaneously accounting for these and other mechanisms, the composition of unemployment has been studied under the umbrella of labor market frictions. A simplified way to account for these frictions has been to assume that companies and job seekers meet at random in the job market. Failure to coordinate these encounters can then be attributed to frictions.

The seminal work of \cite{hall_theory_1979}, \cite{pissarides_job_1979}, and \cite{bowden_existence_1980} paved the way for the application of random matching models in order to integrate frictions into equilibrium models. A reduced way to capture these matching processes is through the aggregate matching function (AMF). In its most typical form, the AMF takes two quantities as inputs --total unemployment and total number of vacancies-- and returns the total number of successful matches. If the AMF produces unsuccessful matches, even when there are more vacancies than unemployed, it means that the labor market has frictions. Like any aggregation, the AMF implicitly assumes certain regularity in the matching process. These assumptions are convenient for mathematical tractability, but they come with the cost of sacrificing structural information about the labor market frictions. For example, if there are clusters of firms that `trap' labor flows, this information would be destroyed by means of aggregation. Of course, whether it is important to consider such clusters depends on the research question. Nevertheless, today's major challenges for labor policy are inherently dynamic and they demand a granular view of labor markets. Hence, we provide a framework to advance our understanding of labor markets in this direction. 

In order to address the limitations of the AMF, numerous models about its micro-foundations have been formulated. In some cases, they rely on theoretical assumptions that are difficult to observe through empirical data. In other cases, micro-foundations can be extremely specific mechanisms that are difficult to extrapolate to more general contexts, or to link to other mechanisms under a common framework (e.g., geographical distance, social networks, skills mismatch, etc.). Furthermore, even when it is possible to account for multiple micro-foundations simultaneously, it does not take much to end up with overly-complicated models. Therefore, developing an overarching framework that accounts for the highly heterogeneous and complex structure of labor market frictions is something desirable from both positive (to understand labor dynamics) and normative (for policy purposes) points of view. In this paper, we propose a new framework to achieve this goal, inspired in micro-level empirical observations on how individuals move from one company to another throughout their careers.

When one thinks about labor market frictions, there are numerous mechanisms that come to mind, for example, social networks, information asymmetries, geographical distance, industrial compatibility, etc. Altogether, these frictions interact and shape the landscape through which individuals flow from one job to an anther, often experiencing unemployment spells in between. Our theory is not about specific frictions and does not propose new ones; instead, it provides a tool to study unemployment while taking into account the complex landscape that emerges from all frictions and their interactions. The main assumption is that the labor market lives on top of a network of firms. This network reveals the pathways that are most likely to be navigated by job seekers, constraining mobility and bounding unemployment to certain locations in the network; something extremely useful for policy purposes. In this network, the presence or absence of an edge represents a categorical relation between two firms, resulting from the frictions that determine the amount of labor mobility between them. More specifically, the absence of an edge means that labor flows between two disconnected firms are unlikely due to high frictions (at least in the short run), while the opposite is expected for connected firms. Together, firms and edges from the \emph{labor flow network} (LFN) of the economy.\footnote{We must clarify that the LFN is assumed to be exogenous. While the reader may wonder about its endogenous nature, this would be of a different time-scale and complexity. In other words, we decide to assume an exogenous LFN in order to study the effect that its topology has on the composition of aggregate unemployment, while leaving the inquiry of its emergence for a different study.} In the same spirit in which the AMF provides an analytic tool to mediate the matching process, the LFN provides a structured object that allows us to analyze labor dynamics in great detail. Another analogy can be drawn from the urns-balls literature \citep{petrongolo_looking_2001}. Here the urns are distributed across individual firms, and jumps between them are restricted by the connections between them. In fact, the classical urns-balls model is a special case of ours; the case in which the network topology is regular (e.g. there is little variation in the number of connections of the nodes). Therefore, the topology of the LFN plays a crucial role in determining unemployment. As we will show in this paper, the empirical topology of the LFN, makes aggregation of unemployment non-trivial. Furthermore, when firms' hiring behavior correlates through the LFN, aggregate unemployment may be significantly higher than expected under the assumption of no frictional structure. Our main finding is that, in the presence of a high labor supply elasticity, the level of aggregate unemployment is strongly dependent on the structure of the LFN. Furthermore, the distribution of unemployed in the economy is directly linked to the specific topology of the LFN. Our framework provides a new way to think about labor dynamics such that the highly heterogeneous structure of all labor market frictions is taken into account to a great degree of detail. This can be extremely useful to study problems where the propagation of shocks and policies shape the speed and reallocation of labor differently, depending on the specific `pockets' of workers and firms affected.

\subsection{Related works}

The idea of limiting job search to groups of firms is not new or uncommon. For example, mismatch models posit that coordination failures between firms and workers are due to frictions that prevent job seekers from freely moving between submarkets. Conventionally, mobility between submarkets is studied by grouping firms into different categories and analyzing the labor flows that take place between such groups. Since the early contribution of \cite{lucas_equilibrium_1974}, multi-sector matching models have offered a variety of ways to think about frictions between submarkets. An example can be found in \cite{shimer_mismatch_2007}, where inter-submarket flows are modeled as a process where workers and jobs are randomly reassigned to any submarket every period. This reassignment originates from an exogenous stochastic process in which movements between any two submarkets are equally likely. Once workers and jobs have been reallocated, matching takes place in each submarket through local AMFs. In contrast, \cite{sahin_mismatch_2014} assume that, provided with information on vacancies, shocks, and efficiencies, workers periodically choose a submarket to move into. Once labor is reallocated, match creation and destruction take place in each submarket. An alternative approach proposed by \cite{herz_structural_2011} assumes that workers can search for vacancies in any submarket and firms can search for workers in the same way. There are costs associated to searching in each submarket. Therefore, matching depends on the optimal decisions of workers and firms about where to search. Other models combine some of these elements in the tradition of Lucas and Prescott \citep{alvarez_search_2011, carrillo-tudela_unemployment_2013, lkhagvasuren_large_2009, kambourov_occupational_2009}. On the other hand, a related strand of research studies submarkets as spatially delimited units (generally cities) \citep{glaeser_wealth_2009, moretti_local_2011, manning_how_2017, kennan_effect_2011}. These models focus on the effect of local shocks when the economy is in spatial equilibrium, which is useful when we know the spatial location of interest. However, as units of aggregation, spatial partitions can be rather arbitrary.

Whether it is for the whole economy or for submarkets, there are a number of problems that arise from viewing matching in aggregate terms, and here we mention a few. First, when an AMF is responsible of pairing up workers and vacancies, it is assumed that all matches are equally likely. This neglects the importance that specific firms have in reallocating labor within a submarket. Second, defining a submarket is an arbitrary choice that might be well suited for a specific problem, but not necessarily for a broader context. Since these classifications are usually built for taxonomic purposes, they are not designed to minimize inter-submarket flows and maximize intra-submarket, which would capture the structural information of labor market frictions. This problem has been pointed out by \cite{jackman_structural_1987} in their classical paper on structural unemployment:

\begin{quote}
... ``there seems no particular reason why unemployed workers should regard themselves as specific to a particular industry, and in practice the unemployed do move between industries reasonably easily." \cite[pg. 19]{jackman_structural_1987}
\end{quote}

Third, aggregation often assumes that any worker from one submarket is equally likely to transition to another submarket. Furthermore, it ignores the fact that only a few firms are responsible for inter-submarket transitions. These firms are crucial to overall labor mobility since they are diffusion outlets or bottlenecks in the process of labor reallocation. Fourth, aggregation `smooths' the search landscape, enabling firm-to-firm flows that are highly unlikely in the short run. In fact, \cite{guerrero_firmtofirm_2015} have shown that the hypothesis of an AMF is rejected as an explanation of empirical firm-to-firm flows, even at the level of submarkets. Using community detection methods for network data, on one hand, \cite{guerrero_employment_2013} show that conventional classifications such as industries and geographical regions poorly capture the clusters of labor that are detected in employer-employee matched micro-data. On the other, \cite{schmutte_free_2014} uses more aggregate data to perform a community detection analysis that reveals four clusters that do not correspond to traditional ways of classifying labor submarkets. Finally, there are, of course, models where mobility decisions between submarkets are heterogeneous and take place between workers of different types \citep{kennan_effect_2011, cortes_where_2015}. This however, does not solve the problem of homogenizing the matching process within groups of firms and omitting the structure of labor flows within a submarket. For these reasons, a framework that does not rely on arbitrary aggregations to define submarkets would represent a significant methodological improvement. \cite{petrongolo_looking_2001} have suggested the use of graph theory as a potential tool to overcome arbitrary aggregations. We take this approach in order to depart from the established notions of submarkets and, instead, look at labor dynamics as random walks on graphs.

\subsection{A network approach}

Our approach is inspired in a local job search mechanism. When a person looks for a job in search of a vacancy, he or she approaches a group of firms that are `accessible' in the short run. Such group is determined by the frictions of the labor market and we assume that it is specific to the firm where this person was last employed. We represent the correspondence between firms and their respective groups of accessible companies through a LFN. In this network, firms are represented by nodes. An edge between enterprises $i$ and $j$ means that frictions are such that $j$ will be accessible to employees of $i$ and vice versa. Therefore, edges have a categorical nature that represents the possibility (or impossibility in their absence) of labor flows between firms. Firm $i$'s edges determine its first neighbors, which are equivalent to the group of accessible firms to someone employed in $i$. We refer to these firms as $i$'s neighbor firms. As a person progresses through his or her career, he or she traverses the economy by taking jobs at the neighbor firms of past employers. This gradual navigation process is fundamentally different from previous approaches because the identity of the firm (i.e., its position in the LFN) matters in order to determine the employment prospects of the job seeker. There is a number reasons why this is important. To mention a few, it allows to study the composition of aggregate unemployment at the level of the firm; it sheds light on the effect of localized shocks and targeted policies; and it exploits the granularity and inter-firm structure captured in employer-employee matched records. By analyzing the steady-state equilibrium, we obtain analytic solutions that inform us about local unemployment, local flows, firm sizes, profits and firm hiring behavior. In addition, this framework provides new micro foundations of stylized facts such as the Beveridge curve and the employer-size premium.

Network theory has been extensively used to study labor markets in the context of information transmission through social networks. The pioneering work of \cite{granovetter_strength_1973} showed the importance that infrequently-used personal contacts have in acquiring non-redundant information about vacancies. Although Granovetter's hypothesis has been challenged by studies that use comprehensive social media micro-data \citep{gee_social_2014,gee_paradox_2014}, the importance of social networks in diffusing job information is not in question. Other studies about the role of social networks in labor markets look at migration \citep{munshi_networks_2003}, urban and rural unemployment \citep{wahba_density_2005}, investment in personal contacts \citep{galeotti_endogenous_2014}, local earnings \citep{schmutte_job_2010}, board interlock networks \citep{kitti_search_2017}, and causality between social connections and hiring decisions \citep{eliason_causal_2012} among other topics. There is also a substantial number of theoretical models about social networks in labor markets and their formation; pioneered by \cite{boorman_combinatiorial_1975} and \cite{montgomery_social_1991}. Some them focus on labor outcomes as a result of the structure of social networks \citep{calvo-armengol_effects_2004, calvo-armengol_job_2005, calvo-armengol_networks_2007, schmutte_job_2010, galenianos_hiring_2014}. Other works analyze inequality and segregation effects in the job market \citep{calvo-armengol_effects_2004, tassier_social_2008}. For a comprehensive review of this literature, we refer to a review elaborated by \cite{ioannides_job_2004}.

Despite the wide application of network methods to study labor markets, most of this work was only focused on the role of social networks in communicating information about vacancies. These studies have important applications in long-term policies such as affirmative action law, but are not so useful for short-term policies such as contingency plans in the presence of shocks. Furthermore, the role of the firm in these models becomes trivial if not absent, which is problematic for policies that aim at incentivizing firms. In fact, little has been done to study labor mobility on networks. To the best of our knowledge, there are only a few studies that analyze labor flows through networks. The idea of national-level highly desegregated LFNs was first introduced by \cite{guerrero_employment_2013} to study firm-to-firm labor flows. For this, the authors use employee-employer matched records from the universe of workers in Finland and a sample from Mexico. They characterize the topology of these LFNs and find that network connectivity is highly correlated with employment growth at the firm level. Using US micro-data, \cite{schmutte_free_2014} constructs job-to-job networks in order to identify four job clusters. Mobility between these clusters is highly frictional and dependent on the business cycle. Both studies find that any clusters identified through community detection methods have little correspondence to standard categorizations such as industrial classification, geographical regions, or occupational groups. The LFN framework provides an alternative way to analyze labor dynamics, while contributing to the use of methods from network science in economics. A closer (and growing) type of literature comes from the studies that use LFNs for different questions related to labor mobility. For example \cite{guerrero_understanding_2016} build a theoretical computational model to study the effect of the LFN topology on unemployment in the presence of shocks. \cite{lopez_network_2015} use the method of random walks on graph to estimate the firm size distribution from information on labor firm-to-firm flows. \cite{mondani_evolving_2017} studies the evolution of LFNs in Stockholm. \cite{tong_diffusing_2017} generalize LFNs to multi-layered graphs. More recently, \cite{park_global_2019} have studied the global LFN through Linked-In data of 500 million individuals in order to characterize geo-industrial clusters. From this burgeoning literature, it is clear that the idea and usage of LFNs, as an alternative to the AMF, has become standard. However, we still lack models that integrate LFNs with micro-economic theory. Hence, our work advances this this front by providing the first model of such type.

Overall, our work complements five strands of literature. First it adds to the family of search and matching models in labor economics by introducing the method of random walks on graphs as a tool to analyze labor mobility and aggregate unemployment. It also pushes the boundaries on how employer-employee matched micro-datasets are used today. Second, it contributes to the field of networks in labor markets by expanding the application of network methods beyond the scope of personal contacts, since social networks are difficult to observe on a large-scale\footnote{Although online social networks provide a rich source of information, they are highly susceptible to biases and multiple factors that incentivize individuals to opt out of this form of communication.}. Since LFNs partially capture labor flows induced by personal contacts (people who worked together may recommend each other in the future), they serve as an additional source of information to study the effect of social networks in the labor market. Third, it complements the literature on micro-foundations of the AMF \citep{butters_equilibrium_1977, hall_theory_1979, pissarides_job_1979, montgomery_social_1991, lang_persistent_1991, blanchard_ranking_1994, coles_understanding_1994, coles_marketplaces_1998, stevens_new_2007, naidu_simple_2007}. Because frictions are captured in the form of a network, there is no need to assume an aggregate matching process. Fourth, it strengthens the growing literature of inter-firm networks \citep{saito_larger_2007, konno_network_2009, atalay_network_2011, acemoglu_network_2012, digiovanni_firms_2014}. By avoiding aggregation into arbitrary submarkets, the LFN approach allows studying firm and labor dynamics jointly. Fifth, it contributes to the study of local labor markets by providing a new way of defining localities at the level of the firm, which should facilitate the study of local shocks and their propagation.

This paper is organized in the following way. \Cref{modelDescription} presents the model in two parts. First, we introduce the problem of the firm, which maximizes profits in the steady-state when wages are exogenous. Then, we characterize the job search process as a random walk on a graph, which helps us solving the firm's problem of choosing the optimal number of vacancies. In \cref{sec:equilibriumUnemployment} we endogenize wages and find that the hiring behavior of the firm correlates with the one of its neighbors. Under an inelastic labor supply, this behavioral correlation becomes systemic, making aggregate unemployment sensitive to the network topology. We illustrate this with hypothetical networks and by performing a counter-factual analysis with empirical data. Finally, in \cref{sec:discussion} we discuss the results, their policy implications and potential of this framework for future research.

\section{Model with an exogenous wage} \label{modelDescription}

The aim of our model is to understand the link between the topology of an exogenous LFN and aggregate unemployment. The model considers an economy in a steady state where firms demand a constant amount of labor and workers search for jobs randomly. Since we are interested in understanding the effect of the network topology on unemployment, we focus on firm behavior and model workers as random walkers. In this section, we assume a single exogenous wage, and in \cref{sec:equilibriumUnemployment} we introduce an exogenous supply to generate wage dispersion.

\subsection{Firms}

There are $\{1,...,N\}$ firms in the economy. In the steady state, each firm has size $L_i$. Every period, a fraction of the $i$'s employees becomes separated with an exogenous probability $\lambda$. By hiring job applicants, the firm compensates the loss of employees. Profits are made exclusively from labor rents. Therefore, firms maximize profits by determining the optimal number of vacancies to open every period. However, opening vacancies also depends on exogenous shocks in the form of investments. These investments enable firms to open vacancies and they arrive with a probability $v$. Therefore, when a firm has vacancies we say that it is \emph{open}, and \emph{closed} otherwise. The expected steady-state firm size is $$(1-\lambda)L_i + vh_iA_i,$$ where $A_i$ is the number of job applicants and $h_i$ is the fraction of applicants hired by the firm.

Unfilled vacancies are destroyed every period, so we use $h_i$ as a continuous approximation for vacancies in a firm. The intuition is that the firm has an expectation about the number of applicants that it would receive in the steady state; for example, by counting the CVs that job seekers drop at its offices everyday. Hence $A_i$ is the expected number of applicants. The firm opens at most $A_i$ positions in order to minimize the cost of unfilled vacancies (of course there still can be unfilled vacancies if the number of applicants is lower than expected). Therefore, the number of vacancies opened by the firm can be written as a fraction of $A_i$. We assume independence between workers, so we can treat $h_i$ as a probability. We call it the \emph{hiring policy}, and it represents the likelihood that a job seeker who applies to firm $i$ becomes hired. 

Building on \cite{barron_employer_1987}, we assume that the objective of the firm is to maximize profits by setting an optimal hiring policy. For this, the firm also takes into account its linear technology with a productivity factor $y$, the exogenous wage $w$, and cost parameters $c \in (0,1)$ and $\kappa \in [0,1]$. Then, the profit maximization problem is given by

\begin{equation}
\max_{h_i} \Pi_i = (1-\lambda )(y-w)L_i + v(y-w)h_iA_i - v cL_ih_i - (1-v) \kappa cL_ih_i.
\label{eq:profitProblem}
\end{equation}

On the one hand, $c$ captures the cost of opening more vacancies. We assume that this cost scales with firm size (e.g., because larger firms invest more in screening processes and HR in general), as suggested by recent empirical evidence \citep{muehlemann_structure_2016}. On the other hand, $\kappa$ represents a sunk cost from HR which the firm incurs when it is closed (e.g., setup expenses for screening future applications). Parameters $c$ and $h$ also affect the sunk cost since the setup cost of HR is assumed to be proportional to the expected number of new hires or vacancies.

In order to generate concavity in \cref{eq:profitProblem}, we assume that the firm understands the job search process for a given set of hiring rates of its neighbors. That is, the firm does not know how other companies arrive to their hiring rates, it only knows the rates of its immediate neighbors and how they determine --among other quantities to be discussed below-- the expected number of incoming job applications.\footnote{This is consistent with the idea of firms having a limited ability to understand the complexity of the system in which they are embedded.\citep{simon_rational_1979}} Therefore, we proceed to characterize job search and obtain the steady-state solutions for $L_i$ and $A_i$, which the firm takes into account in order to maximize profits.

\subsection{Job search \label{sec:jobSearch}}

Let us consider a network where each node represents a firm and the absence of an edge between two firms means that labor market frictions between them are so high that we would not expect any labor flows between them in the short run. This network is represented by the adjacency matrix $\mathbf{A}$, where $\mathbf{A}_{ij}=\mathbf{A}_{ji}=1$ if firms $i$ and $j$ share and edge, and $\mathbf{A}_{ij}=\mathbf{A}_{ji}=0$ otherwise. Workers flow through this network as they gain and lose employment from its nodes, hence the name of labor flow network (LFN). The LFN is unweighted because edges represent a categorical aspect of the labor market: whether we should expect labor flows between two firms or not. It is undirected because the edges capture some `affinity' between firms such that frictions are low in both directions. For simplicity, we do not allow self-loops, so the diagonal entries of $\mathbf{A}$ are all zero. We assume that the LFN has a single component. However, the results are generalizable for networks with multiple components. Firm $i$ has $k_i = \sum_j \mathbf{A}_{ij}$, also known as the degree of $i$. The set $\Gamma_i$ contains all firms $j$ such that $\mathbf{A}_{ij}=1$.

Workers can be in one of two states: \emph{employed} or \emph{unemployed}. Regardless of his or her state, each worker is always associated with a firm. Therefore, jobless workers are associated to their last employers. Each worker employed by firm $i$ might become unemployed with probability $\lambda$. If unemployed, he or she looks at the set $\gamma_i \subseteq \Gamma_i$ of $i$'s neighbor firms that received investments. Hence, we say that $\gamma_i$ is the set of open neighbors of $i$ and it may change from one period to another. If $|\gamma_i|=0$, the job seeker remains unemployed for the rest of the period. Otherwise, he or she selects a firm $j \in \gamma_i$ at random with uniform probability and submits a job application. For simplicity, we assume that each job seeker can submit at most one application per period. It is possible to return to $i$ as long as the last job was held at $j$ such that $\mathbf{A}_{ij}=1$. Finally, if the job application is successful (with probability $h_j$), the job seeker becomes employed at $j$, updating its firm association. Otherwise, it remains unemployed for the rest of the period. We summarize this process in the following steps. 

\begin{enumerate}
    \item Each firm receives an investment with probability $v$.
    \item Each employed worker becomes unemployed with probability $\lambda$.
    \item Each unemployed associated to $i$ (excluding the newly separated ones) picks a firm $j \in \gamma_i$ at random and becomes employed with probability $h_j$.
\end{enumerate}

The reader may be concerned about the possibility that a job seeker may occasionally search among firms that are not connected to his or her last employer. If the probability of such event is low, the model preserves the same characteristics because the LFN induces a dominant effect on job search. When this probability is large, the model becomes an `urns-balls' model, so the structure of the network is irrelevant. What should be the empirically relevant magnitude of such probability? Previous work shows that the idea of searching on a network is empirically compelling since firm-to-firm labor flows tend to be significantly persistent through time \citep{lopez_network_2015}. Other, unrestricted random matching between firms and workers is formally rejected when looking at employer-employee matched records \citep{guerrero_firmtofirm_2015}. These results suggest that, in a more general model, the probability of searching `outside' of the network has to be calibrated with a low value.\footnote{Such a model can be easily constructed, but its solutions do not have an explicit form. In contrast, focusing exclusively on job search `on' the network yields explicit solutions, which is convenient for building economic intuition.}

\subsection{Dynamics}

The stochastic process previously described is a random walk on a graph with waiting times determined by the investment shocks $v$, the separation rate $\lambda$, and the set of hiring policies $\{h_i\}_{i=1}^N$. In order to characterize its dynamics, we concentrate on the evolution of the probability $p_i(t)$ that a worker is employed at firm $i$ in period $t$, and the probability $q_i(t)$ that a worker is unemployed in period $t$ and associated to firm $i$. For this purpose, let us first construct the dynamic equations of both probabilities to then obtain the steady-state solution.

In period $t$, the probability that a worker is employed at firm $i$ depends on the probability $(1-\lambda)p_i(t-1)$ that he or she was employed at the same firm in the previous period and did not become separated. In case that the worker was unemployed during $t-1$, then $p_i(t)$ also depends on: the probability $q_j(t-1)$ that the worker was associated to a neighbor firm $j$; on the probability $\Pr(\gamma_j^{(i)})$ of having a particular configuration $\gamma_j^{(i)}$ of open and closed neighbors of $j$ such that $i$ is open; and on the probability $1/|\gamma_j^{(i)}|$ that the worker picks $i$ from all of $j$'s open neighbors. Altogether, summing over all possible neighbors and all possible configurations of open neighbors, and conditioning to the hiring policy, the probability that a worker is employed by firm $i$ in period $t$ is

\begin{equation}
p_i(t) = (1-\lambda)p_i(t-1) + h_i \sum_{j \in \Gamma_i}q_j(t-1) \sum_{\{\gamma_j^{(i)}\}} \Pr \left(\gamma_j^{(i)} \right) \frac{1}{\left| \gamma_j^{(i)} \right|},
\label{eq:p}
\end{equation}
where $\{\gamma_j^{(i)}\}$ denotes the set of all possible configurations of open and closed neighbors of $j$ where $i$ is open.

The probability that a worker is unemployed during $t$ while associated to firm $i$ depends on the probability $\lambda p_i(t-1)$ of becoming separated from $i$ in the previous period. On the other hand, if the worker was already unemployed, the probability of remaining in such state depends on: the probability $\Pr(\gamma_i=\emptyset)$ that no neighbor firm of $i$ is open and the probability $1-h_j$ of not being hired by the chosen open neighbor $j$. Accounting for all possible non-empty sets $\gamma_i$ of open neighbors, the probability of being unemployed in $t$ and associated to firm $i$ is given by

\begin{equation}
q_i(t) = \lambda p_i(t-1) + q_i(t-1) \left[ \sum_{\gamma_i \neq \emptyset} \Pr(\gamma_i) \frac{1}{\left| \gamma_i \right|} \sum_{j \in \gamma_i} (1-h_j) + \Pr(\gamma_i=\emptyset) \right].
\label{eq:q}
\end{equation}

Up to this point, the model might seem complicated due to all the parameters involved. However, our intention is to provide a general framework that allows the user to control for different degrees of freedom. As we will show ahead, the steady-state solutions take very simple and intuitive forms, while several parameters can be disregarded if no data is available to measure them. Generally speaking, the qualitative nature of our results holds for different calibrations.

\subsection{Steady state}

In the steady-state, $p_i(t)=p_i(t-1)=p_i$ and $q_i(t)=q_i(t-1)=q_i$ for every firm $i$. The following results follow from solving \cref{eq:p,eq:q}.

\begin{prop}
The process specified in \cref{sec:jobSearch} has a unique steady-state where probabilities $p_i$ and $q_i$ are time-invariant for every firm $i$.
\label{prop:steadyState}
\end{prop}
Existence follows from a standard result in random walks on graphs \citep{bollobas_modern_1998} (see appendix). Uniqueness comes from condition $$1=\sum_{i=1}^Np_i + \sum_{i=1}^Nq_i,$$ which indicates that all probabilities should add up to one, implying that every worker is either employed or unemployed, and associated to only one firm. This result implies that a unique steady-state is always reached regardless of how the hiring policies in $\{h_i\}_{i=1}^N$ are assigned to each firm in the LFN. \cite{lopez_network_2015} provide more general results for heterogeneous separation rates and heterogeneous investment shocks. However, this version is more suitable for economic modeling because it yields explicit solutions with intuitive economic meaning.

\begin{prop}
The steady-state average size of a firm $i$ that follows \cref{eq:p,eq:q} is
\begin{equation}
L_i = \frac{\varphi}{\lambda}h_i\bar{h}_{\Gamma_i}k_i,
\label{eq:Lh}
\end{equation}
where $\bar{h}_{\Gamma_i}=\frac{1}{k_i}\sum_j\mathbf{A}_{ij}h_j$ is the average hiring policy of $i$'s neighbor firms and $\varphi$ is a normalizing constant.
\label{prop:firmSize}
\end{prop}

For now, let us defer the explanation of $\varphi$ for a few paragraphs. \Cref{eq:Lh} suggests that, \emph{ceteris paribus}, the size of a firm increases with its degree. As expected, firms can increase their own sizes through larger hiring policies. \Cref{eq:Lh} captures an externality: a firm's hiring policy affects the size of its neighbor firms. This result follows from an intuitive mechanism. If firm $i$ hires more people from its pool of applicants, it increases its own size. In consequence, more people will become separated from $i$ through the exogenous separation process governed by $\lambda$ (which also reduces the size of the firm). More unemployed individuals associated to $i$ translates into a larger pool of job seekers that will potentially apply for a job at $i$'s neighbor $j$. Therefore, if everything else is constant, $A_j$ increases, contributing to $j$'s growth. This mechanism becomes evident in the following result.

\begin{prop}
The steady-state average number of applications received by a firm $i$ that follows \cref{eq:p,eq:q} is
\begin{equation}
A_i = \varphi \bar{h}_{\Gamma_i}k_i.
\label{eq:apps}
\end{equation}
\label{prop:apps}
\end{prop}
The proof follows from the fact that, in the steady-state, the number of separated employees $\lambda L_i$ must equal the number of newly hired ones $h_i A_i$ in order for $L_i$ to remain constant through time (see appendix).

\subsection{Hiring policy and profits}

We assume that firms understand the job search process to a fair extent. That is, they use \cref{eq:Lh,eq:apps} in \cref{eq:profitProblem} and take $\varphi$ and $\bar{h}_{\Gamma_i}$ as given. Then, substituting \cref{eq:Lh,eq:apps} in \cref{eq:profitProblem}, and solving the F.O.C. yields the optimal hiring policy

\begin{equation}
h^* = \frac{\psi}{2 \phi}(y-w),
\label{eq:optH}
\end{equation}
where $\psi = (1 - \lambda +v\lambda)$ and $\phi = c(v + \kappa - v\kappa)$. We have removed sub-index $i$ because the optimal hiring policy is independent of $k_i$. This result is quite intuitive in a neoclassical sense, since higher wages are compensated with lower hiring policies. It also suggests that, with a unique exogenous wage, all firms set the same optimal hiring policy. This means that we can rewrite some of these results exclusively as functions of $k_i$. More specifically, we rewrite the firm size as

\begin{equation}
L_i = \varphi h^{*2} k_i,
\label{eq:Lh2}
\end{equation}
and the profit as

\begin{equation}
\Pi_i^* = \frac{\varphi \psi^3 }{8 \lambda \phi^2}(y-w)^3 k_i.
\label{eq:optProfits}
\end{equation}

\subsection{Aggregation of unemployment\label{sec:unemp}}

Solving \cref{eq:p,eq:q} yields the average number of unemployed individuals associated to firm $i$ in the steady-state. This is a bottom-up construction that takes into account how unemployment is distributed across firms, so we term it \emph{firm-specific unemployment}. This new measure provides information about the employment prospects of a firms' ex-employees and a method to identify pools of local unemployment. Firm-specific unemployment is obtained from the following result.

\begin{prop}
The steady-state average unemployment associated to a firm $i$ that follows \cref{eq:p,eq:q} is 
\begin{equation}
U_i=\frac{\varphi h_ik_i}{1-(1-v)^{k_i}}.
\label{eq:Uh}
\end{equation}
\label{prop:unemp}
\end{prop}

The normalizing constant $\varphi$ captures the population conservation condition $H = \sum_iL_i + \sum_iU_i$, so it takes the form

\begin{equation}
\varphi = \frac{H}{\sum_{i} h_i \bar{h}_{\Gamma_i} k_i \left[ \frac{1}{\lambda} + \frac{1}{\bar{h}_{\Gamma_i} [1-(1-v)^{k_i}]} \right]}.
\label{eq:conservation}
\end{equation}

\Cref{eq:Uh} becomes more intuitive when multiplying by $\frac{\lambda\bar{h}_{\Gamma_i}}{\lambda\bar{h}_{\Gamma_i}}$, in which case we obtain

\begin{equation}
U_i = \frac{\lambda L_i}{\bar{h}_{\Gamma_i} [1-(1-v)^{k_i}]}.
\end{equation}

Note that $\bar{h}_{\Gamma_i} [1-(1-v)^{k_i}]$ is the transition probability from unemployment to employment for a worker associated to firm $i$. The reciprocal of this probability is the average duration  $\bar{t}_i^u$ of an unemployment spell for an individual whose last job was in $i$. Therefore, we can rewrite \cref{eq:Uh} as 

\begin{equation}
U_i = \lambda \bar{t}_i^u L_i
\label{eq:Uh2}
\end{equation}
In general, firm-specific unemployment is an interesting measure because it not only provides a highly granular unit of the composition of aggregate unemployment, but also yields information about how good will be the employment prospects of someone working at a particular company.

Due to the independence between degree and hiring policy implied by \cref{eq:optH}, aggregation of unemployment is straightforward, given that the firm-specific unemployment rate is defined as

\begin{equation}
u_i = \frac{U_i}{U_i+L_i} = \frac{\lambda}{\lambda +h^*[1-(1-v)^{k_i}]},
\label{eq:unempRate}
\end{equation}
which is non-increasing and convex in $k_i$. Note that for a LFN where all firms have the same degree, \cref{eq:unempRate} is equivalent to the Beveridge curve obtained in `urn-balls' models.

Let the LFNs of two economies be represented by their adjacency matrices $\mathbf{A}$ and $\mathbf{A}'$, with corresponding degree distributions $P$ and $P'$, and aggregate unemployment rates $u=\sum_{k=1}^{k_{\max}} u_k P(k)$ and $u'=\sum_{k=1}^{k_{\max}} u_k P'(k)$. Then, the next results follow from network stochastic dominance \citep{jackson_meeting_2007, jackson_relating_2007, lopez-pintado_diffusion_2008}.

\begin{prop}
If $P$ strictly first-order stochastically dominates $P'$, then $u<u'$.
\label{prop:aggU}
\end{prop}

\Cref{prop:aggU} is quite intuitive since the average firm connectivity of $\mathbf{A}$ is higher than in $\mathbf{A}'$. An LFN with higher connectivity reflects an economy with lower labor market frictions. Under these conditions, job seekers have better chances of finding open firms and new job opportunities.

\begin{prop}
If $P'$ is a strict mean-preserving spread of $P$, then $u<u'$.
\label{prop:aggU2}
\end{prop}
Proofs of \cref{prop:aggU,prop:aggU2} follow from direct differentiation of \cref{eq:unempRate}, which shows that $u$ is non-increasing and convex in $k_i$. \Cref{prop:aggU2} means that higher degree heterogeneity translates into more unemployment. Heterogeneity in a LFN reflects the `roughness' of the search landscape. It is analogous to heterogeneity in search and matching models. However, there is the fundamental difference: agents traverse the economy by gradually navigating the LFN, instead of being randomly allocated to any firm. As we will learn ahead, this subtle difference in the reallocation process significantly affects aggregate unemployment when the hiring policies are heterogeneous. We will show that the LFN not only has an ordinal effect on aggregate unemployment, but also a significant impact on its overall level.

\section{Endogenous wages \label{sec:equilibriumUnemployment}} 

Having established that the topology of the LFN affects aggregate unemployment, a natural follow-up question is how this effect works when wages are endogenous. To illustrate this idea, consider the externality through which the hiring rate of a company affects the flows of another. When wages are endogenous, this externality triggers a reaction of the neighbors by updating their hiring policies according to the new expected number of applications (because wages change the costs of sustaining a specific $h$). This, in turn, determines not only firm sizes, but also firm-specific unemployment. Thus, we are interested in studying the set of heterogeneous equilibrium hiring policies $\{h_i^*$\}. For this purpose, we endogenize wages by introducing an aggregate labor supply\footnote{The reader may be inclined for an alternative wage-generating mechanism like Nash bargaining. Although this might be theoretically appealing it would require specifying the worker's behavior and introducing more parameters, in which case, the model becomes intractable. For the purpose of presenting the idea of firms' correlated behaviors through the LFN, it is enough to account for an aggregate supply.}. Equilibrium wages are formed when the individual labor demand of each firm meet the supply, generating a dispersion that depends on the network topology.

For analytic tractability, we adopt a labor supply with a functional form that guarantees a wage bounded by $(0,1)$. However, any other monotonically increasing function can be used, as long as the necessary considerations are made in order to guarantee wages and hiring policies with consistent bounds. The inverse labor supply has the form

\begin{equation}
w_i = \frac{a \ell_i}{b + \ell_i},
\label{eq:laborSupply}
\end{equation}
where $\ell_i$ is the individual demand of firm $i$; $b > 0$ is a parameter that affects the price elasticity; and $a$ provides the upper bound of the wage. We assume $a=y$ for analytical convenience, guaranteeing non-negative rents from labor.

The labor demand of firm $i$ is equivalent to the number of new hires. Firms are wage takers, so their profit-maximization problem remains unchanged. Therefore the labor demand of firm $i$ takes the form 

\begin{equation}
\ell_i = h_i^*A_i.
\label{eq:laborDemand}
\end{equation}

Substituting \cref{eq:laborDemand} in \cref{eq:laborSupply} and using identity \cref{eq:apps} and then \cref{eq:Lh} yields the equilibrium wage

\begin{equation}
w^*_i = \frac{y \varphi h^*_i \bar{h}^*_{\Gamma_i} k_i}{b + \varphi h^*_i \bar{h}^*_{\Gamma_i} k_i} = \frac{y \lambda L_i}{b + \lambda L_i},
\label{eq:eqWage}
\end{equation}
which explicitly shows that larger firms pay higher wages. In other words this result captures the well-known employer size premium \citep{brown_employer_1989, brown_employers_1990}. It also suggests that firms with higher degree pay higher salaries when compared to other firms with the same $h_i$ and $\bar{h}_{\Gamma_i}$.

Substituting \cref{eq:eqWage} in \cref{eq:optH} yields $i$'s equilibrium hiring policy

\begin{equation}
h^*_i = \min \left( 1,  \frac{\phi b - \sqrt{\phi^2 b^2 + \phi\psi\varphi 2by \bar{h}^*_{\Gamma_i}k_i}}{-2\phi \varphi \bar{h}^*_{\Gamma_i}k_i} \right),
\label{eq:optH2}
\end{equation}
where the firm sets either a fraction $h_i^* \geq 0$ or a corner solution where it hires all applicants. Note that \cref{eq:optH2} in its vector form is a continuous map $T:[0,1]^N \rightarrow [0,1]^N$. Therefore, a set $\{h^*_i \}_{i=1}^N$ exists.

\Cref{eq:optH2} captures the interaction between the hiring behavior of firm $i$ (expressed through $h_i$) and the hiring behavior of its neighbors, correlating hiring policies across the LFN in a negative fashion. This has important implications for the reallocation of labor. For example, if a worker leaves a firm with a low hiring policy, his or her employment prospects will be limited to companies with a similar hiring rate. Therefore, escaping this cluster of poor employment prospects takes longer than in a matching process where hiring policies are well-mixed across firms. This has a profound implication on our understanding of local shocks and unemployment traps because the former exacerbate these bottleneck effects, generating unemployment traps. For instance, we know by \cref{eq:Lh} that a higher $k_i$ induces a larger firm size. Then, the negative correlation between $k_i$ and $h_i$ means that a larger proportion of workers (those in the largest firms) are searching for jobs in firms with lower hiring policies (their neighbors). Following this logic, we can expect that an LFN with a degree distribution that is a mean-preserving spread of another one induces a higher level of unemployment.

There is an important connection between the topology of the LFN and the optimal hiring policies. Its importance relies on the degree of heterogeneity of the set $\{h^*_i \}_{i=1}^N$. If there is a large spread of hiring policies, then the effect of the network topology on aggregate unemployment is larger. In this model, the diversity of hiring policies comes from the supply elasticity, since it is the main determinant of wage dispersion\footnote{However, the model is flexible enough to allow firm heterogeneity in parameters such as the separation rate $\lambda$, the productivity $y$, the hiring cost $c$, and the sunk cost $\kappa$. This is an important strength of the model because it facilitates more realistic calibrations that consider the cross-sectional variation of firms.}. \Cref{fig:eqIntuition} illustrates the relationship between supply elasticity, wages, and hiring policies. To build intuition, consider the firm with the largest labor demand $\ell_{\max}$, which determines the maximum wage in the economy. Considering everything else constant, the latter is higher in an economy with a more inelastic labor supply. A higher wage implies a lower hiring policy for this firm, increasing the dispersion between the maximum hiring policy $h_{\max}$ and the lowest one $h_{\min}$. Firms with different degrees set different hiring policies (assuming that $\bar{h}_{\Gamma_i}$ does not cancel the effect of $k_i$). Therefore, heterogeneity in both, wages and the topology of the LFN, are important, so this modeling framework seems adequate and points to important network effects that have not been previously studied.

\begin{center}
\begin{figure}[h!]
\caption{Wage dispersion and hiring policies\label{fig:eqIntuition}}
\begin{centering}
\includegraphics[scale=.6]{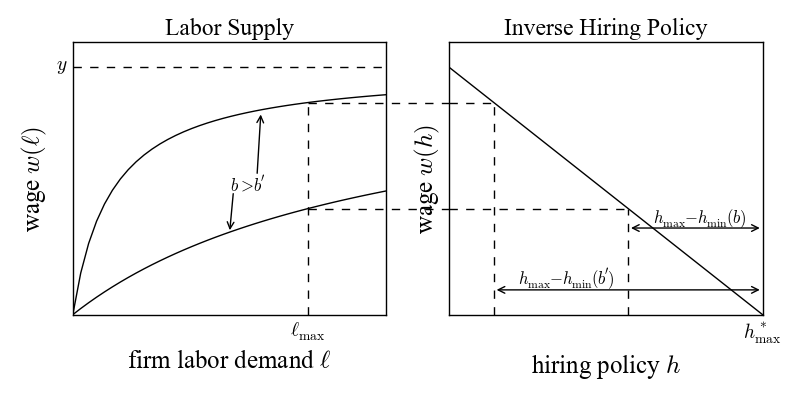}
\par\end{centering}
\footnotesize
The left panel shows two aggregate labor supplies with different elasticities obtained from \cref{eq:laborSupply}. It also presents the corresponding wages that the firm with the largest demand $\ell_{\max}$ would have to pay when confronting the supply. The right panel maps these wages through \cref{eq:optH}, into the hiring policies that would be set by the firm with the largest demand.
\end{figure}
\par\end{center}

In order to better understand the aggregation of unemployment, it is important to analyze how the correlation of hiring policies interact with the topology of the LFN. For this purpose, we study three representative cases of well-known network topologies and then proceed to apply our model to a topology obtained from empirical data.

\subsection{Stylized networks}

We are interested in learning how big are the effects of the LFN topology on aggregate unemployment when hiring policies correlate according to \cref{eq:optH2}. For this purpose, we analyse the outcome of the model under three random networks that relate to homogeneous and heterogeneous job search processes. The first is a regular graph, i.e. a network where every firm has the same number of connections. The second is the popular  Erd\H{o}s-R\'{e}nyi graph, where the firm degree follows a binomial distribution. The third is the so-called scale-free network, which has Pareto-distributed degrees. Naturally, there are other elements that define the topology of a real-world network (e.g., clusters, path length, closeness, etc.). Here, we focus on the degree to build an initial intuition, while we concentrate on empirical LFNs in the \cref{sec:empNets}.

Each of these stylized networks is differentiated by its degree heterogeneity. To study the effect induced by such difference, we have chosen topologies with the same average degree $\bar{k}$, i.e. with a Dirac delta distribution. Therefore, an Erd\H{o}s-R\'{e}nyi degree distribution is a mean-preserving spread of the regular graph, while the scale-free is a mean-preserving spread of the other two. In addition, processes that take place on the regular and the Erd\H{o}s-R\'{e}nyi graphs can be well-approximated by aggregations because the degree heterogeneity is negligible. This is not the case in for the scale-free network, so it is important that we study the differences produced by these three topologies. For the case of the regular graph, it is easy to obtain a closed form solution of \cref{eq:optH2} by substituting $\bar{h}^*_{\Gamma_i}$ by $h^*$ in \cref{eq:optH2} and using \cref{eq:conservation}. This yields

\begin{equation}
h^* = \frac{bN(y\psi \theta - 2\lambda \phi) + \sqrt{b^2N^2(2\lambda\phi+y\psi \theta)^2 + 8byNH\lambda^2 \phi\psi \theta}}
{4\phi\theta(bN+H\lambda)},
\label{eq:optH3}
\end{equation}
where $\theta = 1-(1-v)^k$. For the case of the networks with heterogeneous degrees, the solutions are obtained numerically. A formal proof of the uniqueness of the fixed point in \cref{eq:optH2} is not straightforward. However, numerical experiments via Monte Carlo simulation suggest that \cref{eq:optH2} is a contraction mapping, providing a consistent solution that we use to compute aggregate unemployment.

Panel A in \cref{fig:eqOutcomes} shows the Beveridge curves generated by the model. Here, we portray the Beveridge curve as the relationship between the unemployment rate and the average hiring policy. The curves are generated by solving the model for different levels of the hiring cost $c$ in the interval $[0.1, 0.9]$. There are two notable features that stand out in this diagram. First, the curve from the scale-free network is significantly distant from the other two. Second, the three curves collapse when $\bar{h^*}=1$. This is quite intuitive when we consider the sampling process that workers undergo in the LFN. If all firms set hiring policies near 1, the likelihood of getting a job depends mostly on the investment shocks, which happen uniformly across firms. In this situation, a job seeker at a firm with fewer edges has almost the same chance of finding a job as a worker at a firm with many connections. This also relates to the dispersion of $\{h^*_i \}_{i=1}^N$ because, when firms hire all applicants, there is no diversity of hiring policies, so the LFN effect vanishes.

Panel B in \cref{fig:eqOutcomes} shows the employer-size premium across the three networks. It is clear that the network with largest degree heterogeneity also has the largest wage dispersion. The topology of the network does not shift the $L-w$ curve so we cannot expect significant changes in the average wage due to network structure. Panel C demonstrates the interaction between firms' hiring behavior and their neighbors'. As suggested in \cref{eq:optH2}, there is a negative relationship between $h^*_i$ and $\bar{h}^*_{\Gamma_i}$. These correlations are clustered by levels of $h^*_i$ and their dispersion is larger in the scale-free network.

As shown in panel D of \cref{fig:eqOutcomes}, firms with more edges tend to set lower hiring policies. The mechanism is straightforward: with more neighbors, $A_i$ grows and so does $i$'s demand for labor. More demand implies a higher wage to be paid by the firm, which shifts its profit curve to the left. In order to compensate for higher salaries, the firm needs to re-adjust $h^*_i$ to a lower level. Finally, as predicted by \cref{eq:Lh,eq:Uh}, firms with higher connectivity tend to be larger and have more associated unemployed individuals. In addition, the network with a Pareto degree distribution also exhibits a larger firm size dispersion.

\begin{center}
\begin{figure}[h!]
\caption{Equilibrium outcomes on different network topologies\label{fig:eqOutcomes}}
\begin{centering}
\includegraphics[scale=.6]{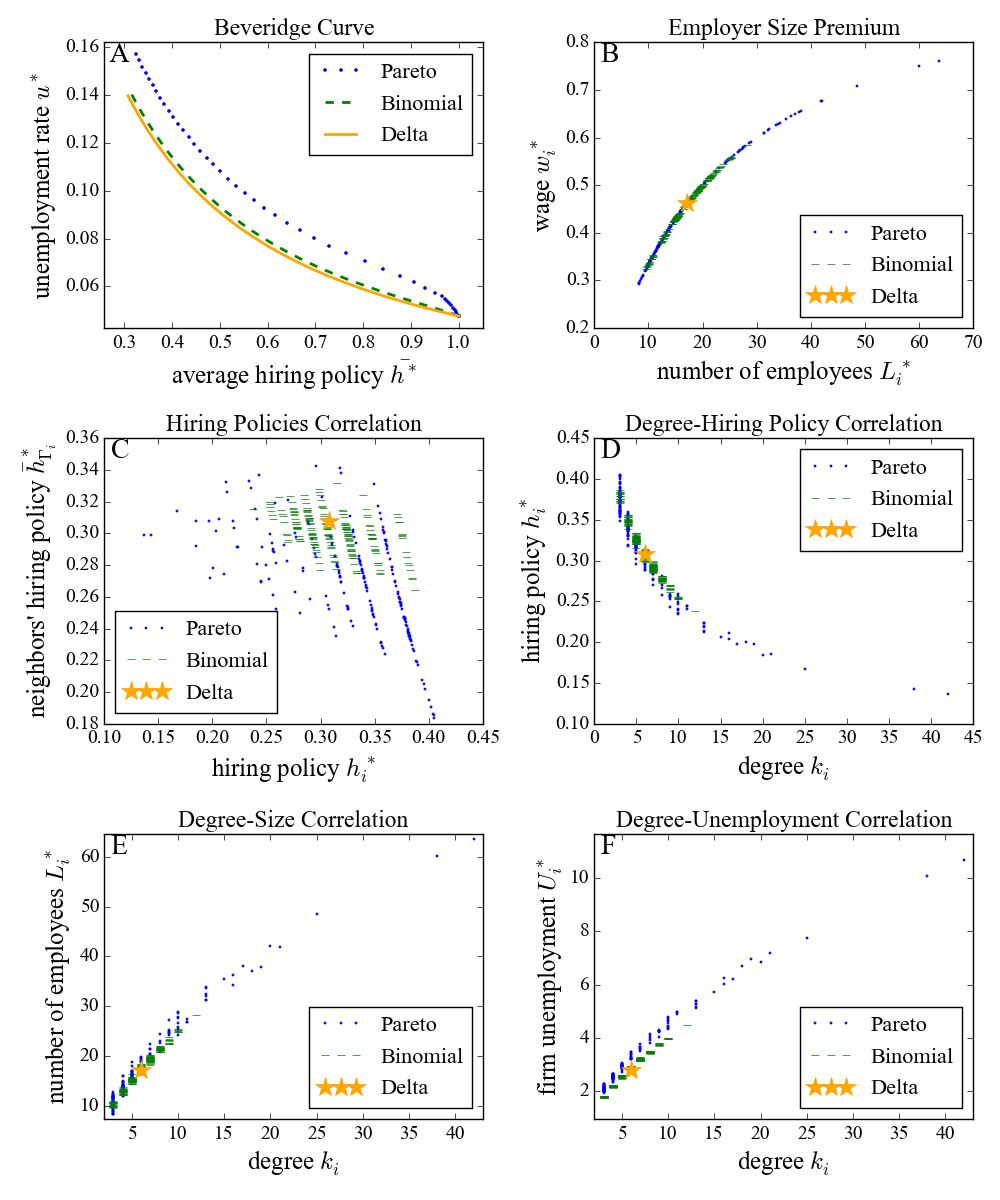}
\par\end{centering}
\footnotesize
Equilibrium solutions for an example calibration: \{$N=200$, $H=4000$, $\lambda=.05$, $y=1$, $v=.8$, $c=.1$, $\kappa=.5$, $b=1$ \}, and different network topologies with the same average degree of 6. The solution for the network with a Dirac delta degree distribution was obtained through \cref{eq:optH3}, while the ones for the binomial and Pareto degree distributions were obtained numerically. Panel a shows the solutions for different levels of $c$. The rest of the panels show the cross-sectional variation of the solution for representative networks.
\end{figure}
\par\end{center}

These results demonstrate that the level and distribution of unemployment are highly sensitive to topologies with significant heterogeneity. This is so because firms' behaviors correlate due to the gradual and restricted movement of labour throughout the economy. Sensitivity to network topology highlights the importance of considering the structure of the labour market frictions, i.e. an empirical LFN. In order to provide an illustration of how these insights could be used empirically, we calibrate the model to a micro-data set and show different counterfactuals on aggregate unemployment that would result from removing the heterogeneous structure of the empirical LFN of Finland.

\subsection{Application \label{sec:empNets}}

We would like to conclude by analyzing real-world LFNs and learning something about the empirical implications of their topologies. So far, most empirical work looking at labor flows is on aggregate datasets (\emph{e.g.}, industries and regions). In contrast, our application exploits firm-level data and demonstrates the importance of understanding the effects of the network topology on aggregate unemployment; something that could be extremely useful for employment policy.

\subsubsection{Data and LFN}

We use the Finnish Longitudinal Employer-Employee Data (FLEED), which consists of an annual panel of employer-employee matched records of the universe of firms and employees in Finland. The panel was constructed by Statistics Finland from social security registries by recording the association between each worker and each firm (enterprise codes, not establishments), at the end of each calendar year. If a worker is not employed, it is not part of the corresponding cross-section. The result is a panel of 20 years that tracks every firm and every employed individual at the end of each year (approximately $2\times10^5$ firms and $2\times10^6$ workers). 

In previous studies, we have constructed LFNs by performing different statistical tests about the significance of flows between firms \citep{guerrero_firmtofirm_2015, lopez_network_2015}, for example, threshold methods and configuration models. Overall, these exercises have shown systematic empirical regularities across different levels of temporal aggregations (\emph{e.g.}, a Pareto degree distribution). For this reason (and because this is an illustrative application), here we take a simpler approach. For a given year, we construct an edge between two firms if we observe labor flows between them.

\subsubsection{Calibration}
Then, we calibrate the model to match the observed aggregate unemployment rates of Finland throughout 20 years, while controlling for its LFNs and separation rates. In order to estimate $\lambda$, we make use of one last theoretical result

\begin{prop}
The steady-state average number of unemployed who become employed after being associated to a firm $i$ that follows \cref{eq:p,eq:q} is 
\begin{equation}
O_i = \varphi h_i \bar{h}_{\Gamma_i} k_i.
\label{eq:Oh}
\end{equation}
\label{prop:outflows}
\end{prop}
The proof follows from the fact that, in the steady-state, $O_i = \lambda L_i$ (see appendix). The intuition is simple: we can consider firm-specific unemployment as a pool of people that is constant through time. The inflows into $U_i$ are $\lambda L_i$ while the outflows are $O_i$. In order for $U_i$ to be constant, the inflows and the outflows must be equal.

Taking advantage of \cref{eq:Oh}, we use the steady-state condition $O_i = \lambda L_i$ in order to estimate the model

\begin{equation}
O_i = \beta_\lambda L_i + \epsilon_i,
\label{eq:OhEst}
\end{equation}
where $\beta_\lambda = \lambda$. We calibrate the model to a daily frequency, so the estimated separation rate becomes $\hat{\beta}^d_\lambda = 1 - (1-\hat{\beta}_\lambda)^{\frac{1}{365}}$ (see appendix).

During the calibration process, we want to avoid trivial solutions such as homogeneous sets of hiring policies. This is so because homogeneity misses important empirical regularities, for example, wage dispersion, heterogeneous firm sizes, and the employer-size premium. We use parameters $c$, $\kappa$, and $b$ for this purpose. As previously discussed, $b$ allows wage dispersion, so an inelastic labor supply is desirable in order to generate heterogeneous hiring policies. Parameter $c$ determines the overall level of $w_i$ and, hence, of $h^*_i$. Finally, $\kappa$ limits the maximum $w_i$ by making the firm more sensitive to the investment shocks, even when it is closed. We normalize $y=1$ and allow $v$ to be a degree of freedom to calibrate the model and match the observed level of aggregate unemployment.

Once calibrated, we use the model to compute a counter-factual. This consists of evaluating the model under a different network structure, while keeping everything else constant. Put it differently, we estimate what would be the aggregate unemployment rate in Finland if the frictions of the labor market would have a homogeneous structure (an implicit assumption in aggregate job search models). In other words, we compute aggregate unemployment when $k_i=k$, which is given by \cref{eq:unempRate}, where $h^*$ corresponds to the solution of the homogeneous case in \cref{eq:optH3}. We perform this exercise for different supply elasticities in order to gain some insights about the minimum and maximum effects of the network topology.

\subsubsection{Results}

\Cref{fig:eqUnempCounter} shows the difference in aggregate unemployment between the fitted model and the counter-factual. We present results for three levels of supply elasticity\footnote{The bump in the counter-factual of 1997 is caused by an anomaly in the data. Due to changes in data administration, 1997 registers a substantial increase in $N$. Most of these firms have $k_i=1$, so the average degree drops nearly 50\% with respect to 1996.}. As previously discussed, a more inelastic labor supply generates more wage dispersion, which contributes to a larger difference in unemployment between the real LFN and the regular network. We interpret this difference as the contribution of the network structure to aggregate unemployment. Under a very elastic labor supply, the contribution is marginal. However, if the supply is highly inelastic, the contribution of the network topology can account for more than 90\% of the unemployment rate. Given that real economies exhibit wage dispersion, the LFN is likely to have a significant effect on aggregate unemployment.

Naturally, any aggregate model (implying a regular network) could also be calibrated to match the empirical level of unemployment. Thus, the counter-factual of a heterogeneous network structure would yield a higher unemployment rate. The important point in this exercise is that, if one would like to predict  unemployment after a change in parameters, it is likely that the aggregate model will underestimate the change in unemployment because the underlying homogeneous structure is less sensitive. Furthermore, the heterogeneous structure of the LFN is observed from empirical microdata on how labor is actually reallocated, something omitted when aggregating the matching process. Therefore, further investigations in the direction of job search on networks would be desirable in order to better understand labor dynamics and the limitations of aggregate approaches.

\begin{center}
\begin{figure}[h!]
\caption{Equilibrium unemployment and counterfactuals}
\begin{centering}
\includegraphics[scale=.55]{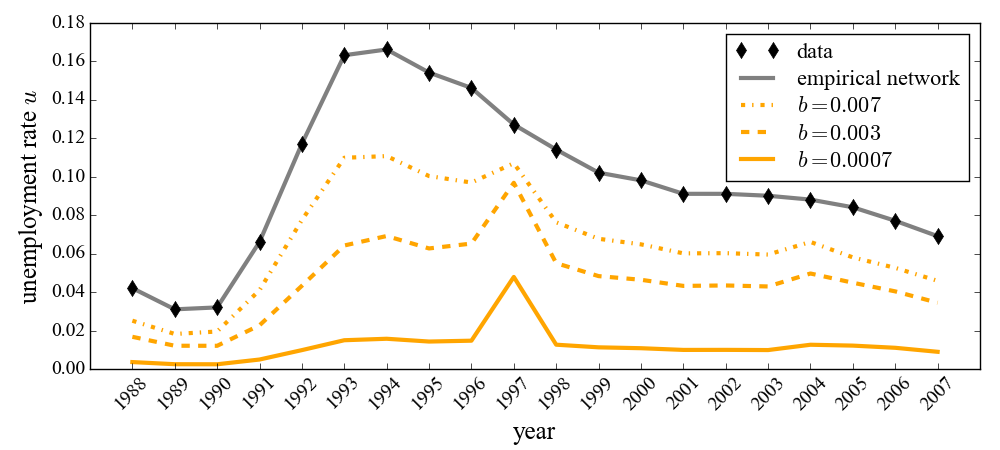}
\label{fig:eqUnempCounter}
\par\end{centering}
\footnotesize
The diamonds correspond to the observed annual aggregate unemployment rate. The grey line was obtained by calibrating the model to match the observed unemployment rates of each year using parameter values: $y=1$, $c=.1$, $\kappa=.5$, and $H=2,000,000$ (the size of the Finnish labor force). $N$ is the number of firms in the data, $\lambda$ was estimated from the data, and $v$ varies between years due to the fitting procedure. 
\end{figure}
\par\end{center}

Finally, the LFN topology not only affects the level of aggregate unemployment, but also its variation through time. In this exercise, it is evident that degree heterogeneity also increases the magnitude of annual variations of the unemployment rate. This is an important result considering that the origins of unemployment volatility is a highly debated topic \citep{mortensen_more_2007, pissarides_unemployment_2009, shimer_labor_2010, obstbaum_finnish_2011}. If structural changes or shocks take place (e.g., changes in $\lambda$ or $v$), the labor reallocation process is smoother on a regular structure than on a heterogeneous one. This is quite intuitive when thinking in terms of job search as a gradual navigation on a network. A shock or a structural change generates heterogeneous adjustments of hiring policies when the network is not regular (and assuming wage dispersion). If the LFN has firms that concentrate many connections, labor reallocation becomes susceptible to the congestion effects that these companies generate by re-adjusting their hiring policies. In a regular topology the reallocation process is smoother because the shock or structural change generates the same re-adjustment across all firms, which happens to have the same number of employees and associated unemployed. Therefore, the LFN points towards the need to understand the propagation of shocks and structural changes through the gradual reallocation of labor that takes place on the network, something that we leave for future work. On a final note, since the reader might be interested in more realistic specifications (e.g. a mixed model with random jumps and multiple choices), we also encourage research into agent-based models, which are far more suitable tools to deal with the complexity of labor markets.

\section{Conclusion} \label{sec:discussion}

In this paper, we develop a theory to study aggregate unemployment from new network-theoretic micro-foundations of job search as a gradual navigation process. The framework allows to study the composition of aggregate unemployment with a resolution at the level of each firm. It also shows that an externality emerges between neighbor firms: \emph{my growth affects yours}. We find that, when labor is reallocated through heterogeneous networks, hiring policies correlate negatively between neighboring firms. For a large network, this becomes systemic, generating wage dispersion in presence of an inelastic labor supply. This dispersion, in turn, causes the level and composition of aggregate unemployment to be dependent on the topology of the LFN. If such topology exhibits high levels of heterogeneity (e.g., a Pareto degree distribution), not only the distribution of firm-specific unemployment is skewed, but the level of aggregate unemployment is significantly higher than expected in a model that neglects the network structure of frictions. This framework provides a rich yet parsimonious description of decentralized labor markets with the possibility of preserving important information that is lost through aggregate approaches.

The LFN framework can be employed to consider firm-specific phenomena. In addition, this framework is particularly well suited to study the propagation of local shocks and structural changes, a major issue in labor policy discussions. Its localized nature allows it to be implemented through other methods such as computer simulation and agent-based models \citep{freeman_war_1998, geanakoplos_getting_2012} in order to study the impact and timing effects of specific policies. This facilitates the study of a richer set of dynamical problems that are difficult to address from an aggregate perspective. For example, we could use employer-employee matched records to calibrate an agent-based model with the real LFN and then simulate local shocks to groups of firms. The computational model would allow obtaining information about how labor would flow out of the affected parts of the economy and, gradually, find its way to firms with better employment prospects. Characterizing this gradual navigation process would be extremely helpful in designing policies that aim not only at alleviating unemployment, but at smoothing transitional phases of the economy.

\newpage
\bibliography{e4}


\pagenumbering{gobble}

\newpage

\appendix

\section{Proof of \cref{prop:steadyState}}

Let $p_i(t)$ and $q_i(t)$ be the probabilities of being employed and unemployed at firm $i$ in period $t$ respectively. Both quantities are dynamically described by

\begin{equation}
p_i(t) = (1-\lambda)p_i(t-1) + h_i \sum_{j \in \Gamma_i}q_j(t-1) \sum_{\{\gamma_j^{(i)}\}} \Pr \left(\gamma_j^{(i)} \right) \frac{1}{\left| \gamma_j^{(i)} \right|},
\label{eq:px}
\end{equation}
and
\begin{equation}
q_i(t) = \lambda p_i(t-1) + q_i(t-1) \left[\sum_{\gamma \neq \emptyset} \Pr(\gamma_i) \frac{1}{\left| \gamma_i \right|} \sum_{j \in \gamma_i} (1-h_j) + \Pr(\gamma_i=\emptyset) + (1-s) \right],
\label{eq:qz}
\end{equation}
where $\gamma_j^{(i)}$ indicates a configuration of open and closed neighbors of $j$, such that $i$ is open. The symbol $\{\gamma_j^{(i)}\}$ denotes the set of all possible configurations of open and closed neighbors of $j$ where $i$ is open. The set $\gamma_i$ contains all open neighbors of $i$, and we denote $\emptyset$ the set of neighbors of $i$ when all of them are closed.

In the steady-state, $p_i(t)=p_i(t-t)=p_i$ and $q_i(t)=q_i(t-t)=q_i$. Note that $\sum_{\gamma \neq \emptyset} \Pr(\gamma_i) + \Pr(\gamma_i=\emptyset) = 1$, so the system defined by \cref{eq:px,eq:qz} becomes

\begin{equation}
0 = -\lambda p_i +h_i \sum_{j \in \Gamma_i} q_j \sum_{\{\gamma_j^{(i)}\}} \Pr \left(\gamma_j^{(i)} \right) \frac{1}{\left| \gamma_j^{(i)} \right|}
\label{eq:zero1}
\end{equation}

\begin{equation}
0 = \lambda p_i - q_i\sum_{\gamma \neq \emptyset} \Pr(\gamma_i) \bar{h}_{\Gamma_i}.
\label{eq:zero2}
\end{equation}
where $\bar{h}_{\Gamma_i}=\frac{1}{k_i}\sum_j\mathbf{A}_{ij}h_j$ is the average hiring policy of $i$'s neighbor firms and $\varphi$ is a normalizing constant.

From \cref{eq:zero2}, let us write $q_i$ in terms of $p_i$ as

\begin{equation}
q_i = \frac{\lambda}{s \sum_{\gamma \neq \emptyset} \Pr(\gamma_i) \bar{h}_{\Gamma_i}} p_i,
\end{equation}
and then substitute $p_i$ with \cref{eq:zero1} to obtain

\begin{equation}
q_i = \sum_{j \in \Gamma_i} \frac{q_j  h_i\sum_{\{\gamma_j^{(i)}\}} \Pr \left(\gamma_j^{(i)} \right)/\left| \gamma_j^{(i)} \right|} {\sum_{\gamma \neq \emptyset} \Pr(\gamma_i) \bar{h}_{\Gamma_i}},
\end{equation}

To understand this further, we write the previous equation in matrix form making use of the adjacency matrix of the graph, $\mathbf{A}$, for which $\mathbf{A}_{ij}=\mathbf{A}_{ji}=1$ if $i$ and $j$ have an edge connecting them, and zero otherwise. This produces the expression

\begin{equation}
\sum_{j=1}^N\left[\mathbf{A}_{ij}\frac{h_i\sum_{\{\gamma_j^{(i)}\}} \Pr \left(\gamma_j^{(i)} \right)/\left| \gamma_j^{(i)} \right|}
{\sum_{\gamma \neq \emptyset} \Pr(\gamma_i) \bar{h}_{\Gamma_i}}-\delta[i,j]\right]\lambda p_j = 0
\end{equation}
for all $i$. This represents a homogeneous system of linear equations, which always has the trivial null solution, and has non-trivial solutions if and only if the matrix contained inside brackets is singular which, among other things, implies that the matrix does not have full rank. 
To show that our model has non-trivial solutions indeed, we define the matrix $\boldsymbol{\Lambda}$, 
with element $\boldsymbol{\Lambda}_{ij}$ corresponding to the expression inside brackets

\begin{equation}
\boldsymbol{\Lambda}_{ij}:=\mathbf{A}_{ij}\frac{h_i\sum_{\{\gamma_j^{(i)}\}} \Pr \left(\gamma_j^{(i)} \right)/\left| \gamma_j^{(i)} \right|}
{\sum_{\gamma \neq \emptyset} \Pr(\gamma_i) \bar{h}_{\Gamma_i}}-\delta[i,j].
\label{eq:Lambda}
\end{equation}
This matrix does not possess full rank as can be explicitly seen from the fact that all columns
add to zero. To show this, we first sum $\boldsymbol{\Lambda}_{ij}$ over $i$

\begin{equation}
\sum_{i=1}^N\boldsymbol{\Lambda}_{ij}=-1+\sum_{i=1}^N \mathbf{A}_{ij}\frac{h_i\sum_{\{\gamma_j^{(i)}\}} \Pr \left(\gamma_j^{(i)} \right)/\left| \gamma_j^{(i)} \right|}
{\sum_{\gamma \neq \emptyset} \Pr(\gamma_i) \bar{h}_{\Gamma_i}}
\label{eq:Lambdasum}
\end{equation}
where $-1$ comes from $-\sum_i\delta[i,j]$. We can now show that the numerator and denominator of the second term are indeed equal. To see this in detail, we organize the elements of $\{\gamma_j^{(i)}\}$ by cardinality $|\gamma_j^{(i)}|$, and rewrite the numerator as

\begin{equation}
\sum_{i=1}^N\mathbf{A}_{ij}h_i\sum_{\{\gamma_j^{(i)}\}}{\rm Pr}(\gamma_j^{(i)})/|\gamma_j^{(i)}|
=\sum_{c=1}^{|\Gamma_j|}\frac{1}{c}\sum_i \mathbf{A}_{ij}h_i\sum_{|\gamma_j^{(i)}|=c}{\rm Pr}(\gamma_j^{(i)}),
\end{equation}
where the last sum is over all elements of $\{\gamma_j^{(i)}\}$ with equal size $c$. Now, the sum over $i$ guarantees
that each neighbor of $j$ belonging to a particular $\gamma_j^{(i)}$ is summed, along with the corresponding $h_r$,
where $r\in\gamma_j^{(i)}$. Therefore, the sum over $i$ can be rewritten as

\begin{equation}
\sum_i \mathbf{A}_{ij}h_i\sum_{|\gamma_j^{(i)}|=c}{\rm Pr}(\gamma_j^{(i)})
=\sum_{|\gamma_j|=c}\left(\sum_{r\in\gamma_j} h_r\right){\rm Pr}(\gamma_j)
\end{equation}
and inserting this into the sum over $c$ leads to

\begin{equation}
\sum_{c=1}^{|\Gamma_j|}\frac{1}{c}\sum_{|\gamma_j|=c}\left(\sum_{r\in\gamma_j} h_r\right){\rm Pr}(\gamma_j)
=\sum_{\gamma_j\neq\emptyset}\frac{\sum_{r\in\gamma_j}h_r}{|\gamma_j|}{\rm Pr}(\gamma_j)
=\sum_{\gamma_j\neq\emptyset}\langle h\rangle_{\gamma_j}{\rm Pr}(\gamma_j)
\end{equation}
Therefore,
\begin{equation}
\sum_{i=1}^N\mathbf{A}_{ij}h_i\sum_{\{\gamma_j^{(i)}\}}{\rm Pr}(\gamma_j^{(i)})/|\gamma_j^{(i)}|
=\sum_{\gamma_j\neq\emptyset}\langle h\rangle_{\gamma_j}{\rm Pr}(\gamma_j)
\end{equation}
which means that for all $j$, \cref{eq:Lambdasum} is identically zero, guaranteeing that the system has non-trivial solutions.

Since the matrix for a connected graph has rank $N-1$, its kernel is one-dimensional, and thus, to choose a unique solution that belongs to the kernel of $\boldsymbol{\Lambda}$ we need a single additional condition. In our case, this condition corresponds to
\begin{equation}
\sum_{i=1}^N(p_i + q_i) = 1,
\label{eq:cond}
\end{equation}
which guarantees that each individual is either employed or unemployed and associated to only one firm each period.  \hfill \emph{Q.E.D.}

\newpage
\section{Proof of \cref{prop:firmSize}}

Let us consider \cref{eq:zero1,eq:zero2} and note that the probability $\Pr(\gamma_i)$ of obtaining a specific configuration $\gamma_i$ of open and closed neighbors follows the binomial $v^{|\gamma_i|}(1-v)^{k_i-|\gamma_i|}$. Then, we obtain that

\begin{equation}
\sum_{\{\gamma_j^{(i)}\}}{\rm Pr}(\gamma_j^{(i)})/|\gamma_j^{(i)}|\rightarrow
\sum_{|\gamma_j^{(i)}|=1}^{k_j}{k_j-1\choose |\gamma_j^{(i)}|-1}\frac{v^{|\gamma_j^{(i)}|}(1-v)^{k_j-|\gamma_j^{(i)}|}}
{|\gamma_j^{(i)}|}
=\frac{1-(1-v)^{k_j}}{k_j}.
\label{eq:simp1}
\end{equation}

For the sum $\sum_{\gamma_j\neq\emptyset}\bar{h}_{\Gamma_i}{\rm Pr}(\gamma_j)$, we note that
each hiring policy $h_i$ for $i\in\Gamma_j$ appears ${k_j-1\choose|\gamma_j|-1}$ times among all the terms 
where there are $|\gamma_j|$ open neighbors to $j$. We can then write

\begin{equation}
\sum_{\gamma_j\neq\emptyset}\bar{h}_{\Gamma_i}{\rm Pr}(\gamma_j)\rightarrow
\sum_{|\gamma_j|=1}^{k_j}{k_j-1\choose|\gamma_j|-1}\frac{\sum_{i\in\Gamma_j} h_i}{|\gamma_j|}
v^{|\gamma_j|}(1-v)^{k_j-|\gamma_j|}
=\bar{h}_{\Gamma_i}(1-(1-v)^{k_j}),
\label{eq:simp2}
\end{equation}
where $\bar{h}_{\Gamma_i}:=\sum_{i\in\Gamma_j}h_i/k_j$, i.e., the average hiring policy of the full neighbor set of $j$. Therefore, \cref{eq:zero1,eq:zero2} simplify into

\begin{equation}
0 = -\lambda p_i +h_i \sum_{i \in \Gamma_i} q_j \frac{1-(1-v)^{k_j}}{k_j}
\label{eq:zero1b}
\end{equation}

\begin{equation}
0 = \lambda p_i - q_i\bar{h}_{\Gamma_i} [1-(1-v)^{k_i}].
\label{eq:zero2b}
\end{equation}

It is easy to see by inspection that the solution to the system is

\begin{equation}
p_i = \frac{\chi h_i \bar{h}_{\Gamma_i} k_i}{\lambda}
\label{eq:solP}
\end{equation}

\begin{equation}
q_i = \frac{\chi h_i k_i}{1-(1-v)^{k_i}}
\label{eq:solQ}
\end{equation}

\begin{equation}
\chi = \frac{1}{\sum_{i} h_i \bar{h}_{\Gamma_i} k_i \left[ \frac{1}{\lambda} + \frac{1}{\bar{h}_{\Gamma_i}[1-(1-v)^{k_i}]} \right]}.
\end{equation}

Given that the workers' actions are independent from each other, the evolution of the firm size follows the binomial 

\begin{equation}
\Pr(L_i) = {{H}\choose{L_i}} p_i^{L_i} (1-p_i)^{H-L_i},
\end{equation}
so the steady-state average firm size $L_i$ (abusing notation) is

\begin{equation}
L_i = Hp_i = \frac{\varphi h_i \bar{h}_{\Gamma_i} k_i}{\lambda}, 
\end{equation}
where $\varphi = H\chi$. \hfill \emph{Q.E.D.}

\newpage
\section{Proof of \cref{prop:apps}}

Consider the probability $a_i(t)$ that a worker submits a job application to firm $i$ in period $t$. This depends on: the probability $q_j(t-1)$ of being unemployed in a neighbor $j \in \Gamma_i$ during the previous period; on the probability $\Pr(\gamma_j^{(i)})$ of $j$ having a configuration $\gamma_j^{(i)}$ of open of closed neighbors in which $i$ is open; and on the probability of choosing $i$ over all other alternative neighbors of $j$. Accounting for all possible events and configurations of neighbors, this probability is written as

\begin{equation}
a_i(t) =\sum_{j \in \Gamma_i} q_j(t-1) \sum_{\{ \gamma_j^{(i)} \}} \Pr\left(\gamma_j^{(i)}\right) \frac{1}{\left| \gamma_j^{(i)} \right|}.
\end{equation}

In the steady-state $a_i(t) = a_i(t-1) = a_i$ and $q_i(t) = q_i(t-1) = q_i$, and by replacing \cref{eq:solQ,eq:simp1} we obtain

\begin{equation}
a_i = \chi  \bar{h}_{\Gamma_i} k_i.
\end{equation}

Since the workers' behaviors are independent from each other, the number of job applications received by firm $i$ in any period follows the binomial

\begin{equation}
\Pr(A_i) = {{H}\choose{A_i}} a_i^{A_i} (1-a_i)^{H-A_i},
\end{equation}
where $H$ is the agent population size, so the steady-state average number of applications $A_i$ (abusing notation) is

\begin{equation}
A_i = Ha_i = \varphi \bar{h}_{\Gamma_i} k_i, 
\end{equation}
where $\varphi = H\chi$. $A_i$ fulfills the steady-state balance condition $\lambda L_i = h_i A_i$. \hfill \emph{Q.E.D.}

\newpage
\section{Proof of \cref{prop:unemp}}

Let us consider the steady-state solution for the probability $q_i$ of being unemployed and associated to firm $i$, as written in \cref{eq:solQ}. Given that the workers' actions are independent from each other, the evolution of the firm-specific unemployment follows the binomial 

\begin{equation}
\Pr(U_i) = {{H}\choose{U_i}} q_i^{U_i} (1-q_i)^{H-U_i},
\end{equation}
so the steady-state average firm-specific unemployment $U_i$ (abusing notation) is

\begin{equation}
U_i = Hq_i = \frac{\varphi h_i k_i}{1-(1-v)^{k_i}}, 
\end{equation}
where $\varphi = H\chi$. \hfill \emph{Q.E.D.}

\newpage
\subsection{Proof of \cref{prop:outflows}}

Consider the probability $o_i(t)$ that a worker associated to firm $i$ finds a job at a different firm in period $t$. This event depends on: the probability $q_i(t-1)$ that the worker was unemployed and associated to firm $i \in \Gamma_j$ during the previous period, on the probability $\Pr(\gamma_i)$ of $i$ having a configuration $\gamma_i$ of open of closed neighbors; and on the probability of choosing one particular firm over all other alternatives available in $\Gamma_i$. Altogether, these factors constitute probability

\begin{equation}
o_i(t) = q_i(t-1) \sum_{ \gamma_i \neq \emptyset} \Pr\left(\gamma_i\right) \frac{1}{\left| \gamma_i \right|}.
\end{equation}

In the steady-state $o_i(t) = o_i(t-1) = o_i$ and $q_i(t) = q_i(t-1) = q_i$, and by replacing \cref{eq:solQ,eq:simp2} we obtain

\begin{equation}
o_i = \chi h_i \bar{h}_{\Gamma_i} k_i.
\end{equation}

Since the workers' behaviors are independent from each other, the number of $i$'s outflows in any period follows the binomial

\begin{equation}
\Pr(O_i) = {{H}\choose{O_i}} o_i^{O_i} (1-o_i)^{H-O_i},
\end{equation}
so the steady-state average outflows $O_i$ (abusing notation) is

\begin{equation}
O_i = Ho_i = \varphi h_i \bar{h}_{\Gamma_i} k_i, 
\end{equation}
where $\varphi = H\chi$. $O_i$ fulfills the steady-state balance condition $O_i = \lambda L_i$. \hfill \emph{Q.E.D.}

\newpage
\subsection{Estimation of separation rates for Finland}

\begin{center}
\begin{table}[ht]
\tiny
\centering
\begin{tabular}{ ccrrcrrcrr }
\hline \hline\\
Year & $\beta_\lambda$ & $N$ & $R^2$ \\ [.6ex]
\hline\\
\multirow{2}{*}{1988} & 0.188*** & \multirow{2}{*}{34,279} & \multirow{2}{*}{0.407}\\  & (2.900e-02)\\ [1ex]
\multirow{2}{*}{1989} & 0.102*** & \multirow{2}{*}{32,771} & \multirow{2}{*}{0.301}\\  & (2.224e-02)\\ [1ex]
\multirow{2}{*}{1990} & 0.105*** & \multirow{2}{*}{25,260} & \multirow{2}{*}{0.246}\\  & (2.442e-02)\\ [1ex]
\multirow{2}{*}{1991} & 0.049*** & \multirow{2}{*}{19,143} & \multirow{2}{*}{0.252}\\  & (9.445e-03)\\ [1ex]
\multirow{2}{*}{1992} & 0.028*** & \multirow{2}{*}{16,810} & \multirow{2}{*}{0.141}\\  & (4.575e-03)\\ [1ex]
\multirow{2}{*}{1993} & 0.188 & \multirow{2}{*}{17,667} & \multirow{2}{*}{0.174}\\  & (1.270e-01)\\ [1ex]
\multirow{2}{*}{1994} & 0.150* & \multirow{2}{*}{20,756} & \multirow{2}{*}{0.279}\\  & (7.513e-02)\\ [1ex]
\multirow{2}{*}{1995} & 0.068*** & \multirow{2}{*}{21,012} & \multirow{2}{*}{0.151}\\  & (1.759e-02)\\ [1ex]
\multirow{2}{*}{1996} & 0.059*** & \multirow{2}{*}{24,076} & \multirow{2}{*}{0.382}\\  & (6.019e-03)\\ [1ex]
\multirow{2}{*}{1997} & 0.065*** & \multirow{2}{*}{51,493} & \multirow{2}{*}{0.509}\\  & (8.652e-03)\\ [1ex]
\multirow{2}{*}{1998} & 0.088*** & \multirow{2}{*}{31,322} & \multirow{2}{*}{0.281}\\  & (1.590e-02)\\ [1ex]
\multirow{2}{*}{1999} & 0.208* & \multirow{2}{*}{33,648} & \multirow{2}{*}{0.409}\\  & (8.141e-02)\\ [1ex]
\multirow{2}{*}{2000} & 0.154** & \multirow{2}{*}{34,008} & \multirow{2}{*}{0.342}\\  & (4.993e-02)\\ [1ex]
\multirow{2}{*}{2001} & 0.088*** & \multirow{2}{*}{33,331} & \multirow{2}{*}{0.323}\\  & (1.667e-02)\\ [1ex]
\multirow{2}{*}{2002} & 0.066*** & \multirow{2}{*}{33,031} & \multirow{2}{*}{0.376}\\  & (1.103e-02)\\ [1ex]
\multirow{2}{*}{2003} & 0.070*** & \multirow{2}{*}{33,842} & \multirow{2}{*}{0.367}\\  & (1.041e-02)\\ [1ex]
\multirow{2}{*}{2004} & 0.592*** & \multirow{2}{*}{35,924} & \multirow{2}{*}{0.609}\\  & (1.528e-01)\\ [1ex]
\multirow{2}{*}{2005} & 0.074*** & \multirow{2}{*}{41,978} & \multirow{2}{*}{0.415}\\  & (1.029e-02)\\ [1ex]
\multirow{2}{*}{2006} & 0.127*** & \multirow{2}{*}{44,403} & \multirow{2}{*}{0.524}\\  & (3.286e-02)\\ [1ex]
\multirow{2}{*}{2007} & 0.086*** & \multirow{2}{*}{42,767} & \multirow{2}{*}{0.470}\\  & (1.171e-02)\\ [1ex]
\hline
\end{tabular}
\caption{Estimation of annual separation rates for Finland via \cref{eq:OhEst}. Robust standard errors in parentheses. * $p<0.05$, ** $p<0.01$, *** $p<0.001$.}
\end{table}
\end{center}


\end{document}